\documentclass[12pt]{article}
\usepackage{amsmath,amsfonts,amssymb}  

\makeatletter
\def\numberbysection{\@addtoreset{equation}{section}
    \def\theequation{\thesection.\arabic{equation}}}
\makeatother

\numberbysection

\newcommand{\be}{\begin{eqnarray}}
\newcommand{\ee}{\end{eqnarray}}
\newcommand{\non}{\nonumber}
\newcommand{\id}{\mathbb{I}}

\newcommand{\diag}{\mathop{\rm diag}\nolimits}

\def\bJ{{\mathbb J}}
\def\bR{{\mathbb R}}
\def\bL{{\mathbb L}}
\def\bQ{{\mathbb Q}}
\def\bC{{\mathbb C}}
\def\bH{{\mathbb H}}

\begin{document}

\begin{titlepage}
\strut\hfill UMTG--266
\vspace{.5in}
\begin{center}

\LARGE Yangian symmetry and bound states\\
in AdS/CFT boundary scattering\\
\vspace{1in}
\large Changrim Ahn \footnote{
       Department of Physics, Ewha Womans University, 
       Seoul 120-750, South Korea; ahn@ewha.ac.kr} and
       Rafael I. Nepomechie \footnote{
       Physics Department, P.O. Box 248046, University of Miami,
       Coral Gables, FL 33124 USA; nepomechie@physics.miami.edu}\\

\end{center}

\vspace{.5in}

\begin{abstract}
We consider the problem of boundary scattering for $Y=0$ maximal giant
graviton branes.  We show that the boundary $S$-matrix for the
fundamental excitations has a Yangian symmetry.  We then exploit this
symmetry to determine the boundary $S$-matrix for two-particle bound
states.  We verify that this boundary $S$-matrix satisfies the
boundary Yang-Baxter equations.
\end{abstract}

\end{titlepage}

\setcounter{footnote}{0}

\section{Introduction}\label{sec:intro}

The AdS/CFT bulk $S$-matrix \cite{St}-\cite{AFZ} has a remarkable
Yangian symmetry $Y(su(2|2))$ \cite{Be2}.\footnote{For earlier
investigations of Yangian symmetry in AdS/CFT, see for example
\cite{DNW}-\cite{Zw}.} Since this $S$-matrix (which is for the
fundamental excitations) can already be determined\footnote{We do not
concern ourselves here with overall scalar factors.  Also, we focus on
just one copy of the $su(2|2)$ $S$-matrix; the full AdS/CFT $S$-matrix
is a tensor product of two such copies.} from just the (centrally
extended) $su(2|2)$ symmetry, the further Yangian symmetry may appear
to be only a mathematical curiosity.  However, in order to construct
higher dimensional (``bound state'') $S$-matrices, the $su(2|2)$
symmetry does not suffice \cite{AF}; and the Yangian symmetry can be
used to completely determine the two-particle \cite{AF, dL} and
general $l$-particle bound state bulk $S$-matrices \cite{AdLT}.  It is
fortunate that such a general way of generating higher-dimensional
$S$-matrices has been found, since the fusion procedure \cite{KRS}
(which has played a very important role for conventional $S$-matrices)
does not seem to work for AdS/CFT $S$-matrices.  Knowledge of
higher-dimensional $S$-matrices is necessary for deriving functional
relations for the corresponding transfer matrices, thermodynamic Bethe
ansatz equations, etc.

We initiate in this note an investigation of higher-dimensional
AdS/CFT {\it boundary} $S$-matrices.  For simplicity, we restrict to
the case of open strings attached to so-called $Y=0$ maximal giant
graviton branes \cite{HM}.  The fundamental boundary $S$-matrix can be
determined from $su(1|2)$ symmetry \cite{HM, AN}.  However, similarly
to the bulk case, we find that this symmetry does not suffice to
determine the bound state boundary $S$-matrices.  We show that the
fundamental boundary $S$-matrix has a residual Yangian symmetry.  We
then exploit this Yangian symmetry to determine the two-particle bound
state boundary $S$-matrix.  In contrast to the fundamental case, this
boundary $S$-matrix is not diagonal.  We verify that it satisfies the
boundary Yang-Baxter equations.  We formulate the discussion in terms
of Zamolodchikov-Faddeev (ZF) operators \cite{ZZ, Fa}, which provide a
convenient way of handling the intricate coproducts of the Yangian
generators.

The outline of this paper is as follows.  In Section
\ref{sec:fundamental} we consider the case of the fundamental
excitations.  After briefly reviewing the $su(2|2)$ and Yangian
symmetries of the bulk $S$-matrix, we show that the boundary
$S$-matrix also has Yangian symmetry.  In Section \ref{sec:boundstate}
we consider the case of two-particle bound states.  After another
brief review of the bulk symmetries in this higher-dimensional
representation, we use the $su(1|2)$ and Yangian symmetries to
determine the boundary $S$-matrix.  We conclude with a brief
discussion of our results.

\section{Fundamental representation}\label{sec:fundamental}

This section is devoted to the scattering of the fundamental 
excitations. We begin with a review of bulk scattering, and then turn 
to boundary scattering.

\subsection{Bulk scattering}\label{subsec:fundamentalBulk}

It is convenient to work with ZF operators.
Following \cite{AFZ}, we denote the fundamental ZF operators by
$A_{i}^{\dagger}(p)$, $i=1\,, 2\,, 3\,, 4$.
The operators with $i = 1, 2$ are bosonic, while the operators with
$i=3, 4$ are fermionic.
These operators create asymptotic particle states of momentum $p$ when
acting on the vacuum state $|0\rangle$. 
The matrix elements of the fundamental bulk $S$-matrix $S^{AA}(p_{1},
p_{2})$ are defined by the relation
\be
A_{i}^{\dagger}(p_{1})\, A_{j}^{\dagger}(p_{2}) = 
S_{\ \ \ \ i\, j}^{AA\ i' j'}(p_{1}, p_{2})\, 
A_{j'}^{\dagger}(p_{2})\, A_{i'}^{\dagger}(p_{1}) \,,
\label{bulkS1}
\ee
where summation over repeated indices is always understood. 

The centrally extended $su(2|2)$ algebra consists of the rotation
generators $\bL_{a}^{\ b}$, $\bR_{\alpha}^{\ \beta}$, the supersymmetry
generators $\bQ_{\alpha}^{\ a}$, $\bQ_{a}^{\dagger \alpha}$, and the
central elements $\bC\,, \bC^{\dagger}\,, \bH$.  Latin indices $a\,, b\,, 
\ldots$ take values $\{1\,, 2\}$, while Greek indices $\alpha\,, 
\beta\,, \ldots$ take values $\{3\,, 4\}$. These generators have the
following nontrivial commutation relations \cite{Be1, AFZ}
\be
\left[ \bL_{a}^{\ b}\,, \bJ_{c} \right] &=& \delta_{c}^{b} \bJ_{a} - 
\frac{1}{2} \delta_{a}^{b} \bJ_{c}\,, \quad 
\left[ \bR_{\alpha}^{\ \beta}\,, \bJ_{\gamma} \right] =
\delta_{\gamma}^{\beta} \bJ_{\alpha} - 
\frac{1}{2} \delta_{\alpha}^{\beta} \bJ_{\gamma}\,, \non  \\
\left[ \bL_{a}^{\ b}\,, \bJ^{c} \right] &=& -\delta_{a}^{c} \bJ^{b} + 
\frac{1}{2} \delta_{a}^{b} \bJ^{c}\,, \quad 
\left[ \bR_{\alpha}^{\ \beta}\,, \bJ^{\gamma} \right] =
-\delta_{\alpha}^{\gamma} \bJ^{\beta} +
\frac{1}{2} \delta_{\alpha}^{\beta} \bJ^{\gamma}\,, \non \\
\Big\{\bQ_{\alpha}^{\ a}\,, \bQ_{\beta}^{\ b}\Big\}&=& 
\epsilon_{\alpha \beta}\epsilon^{a b} \bC \,, \quad 
\Big\{\bQ_{a}^{\dagger \alpha}\,, \bQ_{b}^{\dagger \beta} \Big\}=
\epsilon^{\alpha \beta}\epsilon_{a b} \bC^{\dagger} \,, \non \\
\Big\{\bQ_{\alpha}^{\ a}\,, \bQ_{b}^{\dagger \beta} \Big\} &=& \delta_{b}^{a} 
\bR_{\alpha}^{\ \beta}+ \delta_{\alpha}^{\beta} \bL_{b}^{\ a} 
+ \frac{1}{2} \delta_{b}^{a} \delta_{\alpha}^{\beta} \bH \,,
\label{symmetryalgebra}
\ee
where $\bJ_{i}$ ($\bJ^{i}$) denotes any lower (upper) index of a generator,
respectively. 

The action of the bosonic generators $\bL_{a}^{\ b}\,, \bR_{\alpha}^{\ 
\beta}$ on the ZF operators is given by 
\be
\bL_{a}^{\ b}\, A_{c}^{\dagger}(p) &=& (\delta_{c}^{b}\delta_{a}^{d} - 
\frac{1}{2}\delta_{a}^{b}\delta_{c}^{d}) A_{d}^{\dagger}(p) +
A_{c}^{\dagger}(p)\, \bL_{a}^{\ b} \,, \quad 
\bL_{a}^{\ b}\, A_{\gamma}^{\dagger}(p) = A_{\gamma}^{\dagger}(p)\, 
\bL_{a}^{\ b}\,, \non \\
\bR_{\alpha}^{\ \beta}\, A_{\gamma}^{\dagger}(p) &=& 
(\delta_{\gamma}^{\beta}\delta_{\alpha}^{\delta} - 
\frac{1}{2}\delta_{\alpha}^{\beta}\delta_{\gamma}^{\delta}) 
A_{\delta}^{\dagger}(p) +
A_{\gamma}^{\dagger}(p)\, \bR_{\alpha}^{\ \beta} \,, \quad
\bR_{\alpha}^{\ \beta}\, A_{c}^{\dagger}(p) =  A_{c}^{\dagger}(p)\, 
\bR_{\alpha}^{\ \beta}\,.
\label{repBulk1}
\ee
Moreover, the action of the supersymmetry generators on the ZF 
operators  is given by (see Eq.  (4.21) in \cite{AFZ} and \cite{AN})
\be
\bQ_{\alpha}^{\ a}\, A_{b}^{\dagger}(p) &=& e^{- i p/2} \left[
a(p) \delta_{b}^{a} A_{\alpha}^{\dagger}(p) + 
A_{b}^{\dagger}(p)\, \bQ_{\alpha}^{\ a} \right] \,, \non \\
\bQ_{\alpha}^{\ a}\, A_{\beta}^{\dagger}(p) &=& e^{- i p/2} \left[ 
b(p) \epsilon_{\alpha \beta}\epsilon^{a b} A_{b}^{\dagger}(p) -
A_{\beta}^{\dagger}(p)\, \bQ_{\alpha}^{\ a} \right]\,, \non \\
\bQ_{a}^{\dagger \alpha}\, A_{b}^{\dagger}(p) &=& e^{i p/2} \left[
c(p) \epsilon_{a b} \epsilon^{\alpha \beta} A_{\beta}^{\dagger}(p) + 
A_{b}^{\dagger}(p)\, \bQ_{a}^{\dagger \alpha} \right] \,, \non \\
\bQ_{a}^{\dagger \alpha}\, A_{\beta}^{\dagger}(p) &=& e^{i p/2} \left[
d(p) \delta_{\beta}^{\alpha} A_{a}^{\dagger}(p) - 
A_{\beta}^{\dagger}(p)\, \bQ_{a}^{\dagger \alpha} \right] \,.
\label{repBulk2}
\ee 
It follows that the action of the central charges on the ZF operators is given by
\be
\bC\, A_{i}^{\dagger}(p) &=& e^{-i p}\left[
a(p) b(p) A_{i}^{\dagger}(p) + 
A_{i}^{\dagger}(p)\, \bC \right]\,, \non \\
\bC^{\dagger}\, A_{i}^{\dagger}(p) &=& e^{i p}\left[ 
c(p) d(p) A_{i}^{\dagger}(p) + 
A_{i}^{\dagger}(p)\, \bC^{\dagger} \right] \,, \non \\
\bH\, A_{i}^{\dagger}(p) &=& \left[a(p) d(p) + b(p) c(p)\right] A_{i}^{\dagger}(p) + 
A_{i}^{\dagger}(p)\, \bH \,.
\label{repBulk3}
\ee 
Arutyunov-Frolov-Zamaklar work with a different set of relations for
the supersymmetry generators which involve the world-sheet momentum
operator (see Eq.  (4.15) in \cite{AFZ}).  However, as noted in
\cite{AN}, the relations (\ref{repBulk2}) are more natural when
dealing with a boundary.

The ZF operators form a representation of the symmetry algebra 
provided $a d - b c = 1$.  The parameters can be chosen as follows 
\cite{Be1, AF}
\be
a = \sqrt{\frac{g}{2l}}\eta\,, \quad 
b = \sqrt{\frac{g}{2l}}\frac{i}{\eta}\left(\frac{x^{+}}{x^{-}}-1\right)\,, \quad 
c= -\sqrt{\frac{g}{2l}}\frac{\eta}{x^{+}}\,, \quad 
d=\sqrt{\frac{g}{2l}}\frac{x^{+}}{i \eta}\left(1 - \frac{x^{-}}{x^{+}}\right)\,,
\label{BulkParameters}
\ee
where
\be
x^{+}+\frac{1}{x^{+}}-x^{-}-\frac{1}{x^{-}} = \frac{2l i}{g}\,, \quad 
\frac{x^{+}}{x^{-}} = e^{i p} \,,
\label{xpm}
\ee 
and $l=1$ for the fundamental case under consideration in this section. 
Moreover, following \cite{AF}, we take
\be
\eta = e^{i p/4}\sqrt{i(x^{-}-x^{+})}  \,,
\label{eta}
\ee 
which has an extra factor  $e^{i p/4}$ compared with the 
corresponding quantity in \cite{AFZ, AN}.

The fundamental bulk $S$-matrix (\ref{bulkS1}) is determined by this
$su(2|2)$ symmetry \cite{Be1, AFZ}.  For convenience, we reproduce
here the result for the nonzero matrix elements: \footnote{In order to 
streamline the notation, here we denote the $S$-matrix 
element $S_{\ \ \ \ i\, j}^{AA\ i' j'}(p_{1}, p_{2})$ simply by
$S_{i\, j}^{i' j'}$; i.e., we drop both the $AA$ label and the 
arguments $(p_{1}, p_{2})$.}
\be
S_{a\, a}^{a\, a} &=& {\cal A}\,, \quad 
S_{\alpha\, \alpha}^{\alpha\, \alpha} = {\cal D}\,, \non \\
S_{a\, b}^{a\, b} &=& \frac{1}{2}({\cal A}-{\cal B})\,, \quad 
S_{a\, b}^{b\, a} = \frac{1}{2}({\cal A}+{\cal B}) \,, \non \\
S_{\alpha\, \beta}^{\alpha\, \beta} &=& \frac{1}{2}({\cal D}-{\cal E})\,, \quad 
S_{\alpha\, \beta}^{\beta\, \alpha} = \frac{1}{2}({\cal D}+{\cal E}) \,, \non \\
S_{a\, b}^{\alpha\, \beta} &=& 
-\frac{1}{2}\epsilon_{a b}\epsilon^{\alpha \beta}\, {\cal C} \,, \quad
S_{\alpha\, \beta}^{a\, b} = 
-\frac{1}{2}\epsilon^{a b}\epsilon_{\alpha \beta}\, {\cal F} \,, \non \\
S_{a\, \alpha}^{a\, \alpha} &=& {\cal G}\,, \quad 
S_{a\, \alpha}^{\alpha\, a} = {\cal H} \,, \quad 
S_{\alpha\, a}^{a\, \alpha} = {\cal K}\,, \quad 
S_{\alpha\, a}^{\alpha\, a} = {\cal L} \,,  
\label{bulkS3}
\ee
where $a\,, b \in \{1\,, 2\}$ with $a \ne b$;  
$\alpha\,, \beta \in \{3\,, 4\}$ with $\alpha \ne \beta$; and
\be
{\cal A} &=& S_{0}\frac{x^{-}_{2}-x^{+}_{1}}{x^{+}_{2}-x^{-}_{1}}
\frac{\eta_{1}\eta_{2}}{\tilde\eta_{1}\tilde\eta_{2}} \,, \non \\
{\cal B} &=&-S_{0}\left[\frac{x^{-}_{2}-x^{+}_{1}}{x^{+}_{2}-x^{-}_{1}}+
2\frac{(x^{-}_{1}-x^{+}_{1})(x^{-}_{2}-x^{+}_{2})(x^{-}_{2}+x^{+}_{1})}
{(x^{-}_{1}-x^{+}_{2})(x^{-}_{1}x^{-}_{2}-x^{+}_{1}x^{+}_{2})}\right]
\frac{\eta_{1}\eta_{2}}{\tilde\eta_{1}\tilde\eta_{2}}\,, \non \\
{\cal C} &=& S_{0}\frac{2i x^{-}_{1} x^{-}_{2}(x^{+}_{1}-x^{+}_{2}) \eta_{1} \eta_{2}}
{x^{+}_{1} x^{+}_{2}(x^{-}_{1}-x^{+}_{2})(1 - x^{-}_{1} x^{-}_{2})} 
\,, \qquad
{\cal D} = -S_{0}\,, \non \\
{\cal E} &=&S_{0}\left[1-2\frac{(x^{-}_{1}-x^{+}_{1})(x^{-}_{2}-x^{+}_{2})
(x^{-}_{1}+x^{+}_{2})}
{(x^{-}_{1}-x^{+}_{2})(x^{-}_{1} x^{-}_{2}-x^{+}_{1} 
x^{+}_{2})}\right]\,, \non \\
{\cal F} &=& S_{0}\frac{2i(x^{-}_{1}-x^{+}_{1})(x^{-}_{2}-x^{+}_{2})(x^{+}_{1}-x^{+}_{2})}
{(x^{-}_{1}-x^{+}_{2})(1-x^{-}_{1} x^{-}_{2})\tilde\eta_{1} \tilde\eta_{2}}\,, 
\non \\
{\cal G} &=&S_{0}\frac{(x^{-}_{2}-x^{-}_{1})}{(x^{+}_{2}-x^{-}_{1})}
\frac{\eta_{1}}{\tilde\eta_{1}}\,, \qquad 
{\cal H} =S_{0}\frac{(x^{+}_{2}-x^{-}_{2})}{(x^{-}_{1}-x^{+}_{2})}
\frac{\eta_{1}}{\tilde\eta_{2}}\,, \non \\
{\cal K} &=&S_{0}\frac{(x^{+}_{1}-x^{-}_{1})}{(x^{-}_{1}-x^{+}_{2})}
\frac{\eta_{2}}{\tilde\eta_{1}}\,, \qquad 
{\cal L} = S_{0}\frac{(x^{+}_{1}-x^{+}_{2})}{(x^{-}_{1}-x^{+}_{2})}
\frac{\eta_{2}}{\tilde\eta_{2}}\,, 
\label{bulkS4}
\ee
where 
\be
x^{\pm}_{i} = x^{\pm}(p_{i})\,, \quad 
\eta_{1} = \eta(p_{1}) e^{i p_{2}/2}\,, \quad \eta_{2}=\eta(p_{2})\,, 
\quad \tilde\eta_{1} =\eta(p_{1})\,, \quad \tilde\eta_{2} 
=\eta(p_{2})e^{i p_{1}/2}\,,
\ee
and $\eta(p)$ is given in (\ref{eta}).  This $S$-matrix satisfies the
standard Yang-Baxter equation.

Following \cite{Be2, AdLT}, for each $su(2|2)$ generator $\bJ$, we 
denote the corresponding Yangian $Y(su(2|2))$ generator (in the 
evaluation representation) by
\be
\hat \bJ = -\frac{1}{2}i g u\, \bJ \,,
\ee
where 
\be
u = \frac{1}{2}\left(x^{+}+\frac{1}{x^{+}}+x^{-}+\frac{1}{x^{-}} 
\right) \,.
\ee

The action of the Yangian generators on the ZF operators can be 
inferred from the coproducts given in \cite{Be2, AdLT} and the 
relations (\ref{repBulk1})-(\ref{repBulk3}). For example, from the 
coproduct for $\hat \bL_{2}^{\ 1}$ 
\be
\Delta(\hat\bL_{2}^{\ 1}) &=& \hat \bL_{2}^{\ 1} \otimes \id
+ \id \otimes  \hat \bL_{2}^{\ 1} 
+ \frac{1}{2} \bL_{2}^{\ c}  \otimes \bL_{c}^{\ 1} 
- \frac{1}{2} \bL_{c}^{\ 1}  \otimes \bL_{2}^{\ c} \non \\
& &- \frac{1}{2} \bQ_{2}^{\dagger \gamma}  \otimes \bQ_{\gamma}^{\ 1}
- \frac{1}{2} \bQ_{\gamma}^{\ 1}  \otimes \bQ_{2}^{\dagger \gamma} \,,
\label{coproduct}
\ee
we obtain the relations (which we shall use later)
\be
\hat \bL_{2}^{\ 1}\, A_{1}^{\dagger}(p) &=& -\frac{1}{2}i g u\, 
A_{2}^{\dagger}(p) +  A_{1}^{\dagger}(p)\, \hat \bL_{2}^{\ 1}
-\frac{1}{2} A_{1}^{\dagger}(p)\,  \bL_{2}^{\ 1} 
+ \frac{1}{2} A_{2}^{\dagger}(p)\,  (\bL_{1}^{\ 1}-\bL_{2}^{\ 2}) 
\non \\
&+&\frac{1}{2} c(p) A_{4}^{\dagger}(p)\,  \bQ_{3}^{\ 1} 
-\frac{1}{2} c(p) A_{3}^{\dagger}(p)\,  \bQ_{4}^{\ 1}
-\frac{1}{2} a(p) A_{3}^{\dagger}(p)\,  \bQ_{2}^{\dagger 3}
-\frac{1}{2} a(p) A_{4}^{\dagger}(p)\,  \bQ_{2}^{\dagger 4} \,, \non\\
\hat \bL_{2}^{\ 1}\, A_{2}^{\dagger}(p) &=&
A_{2}^{\dagger}(p)\, \hat \bL_{2}^{\ 1}
+ \frac{1}{2} A_{2}^{\dagger}(p)\,  \bL_{2}^{\ 1} \,, \non\\
\hat \bL_{2}^{\ 1}\, A_{3}^{\dagger}(p) &=&
A_{3}^{\dagger}(p)\, \hat \bL_{2}^{\ 1}
+ \frac{1}{2} d(p) A_{2}^{\dagger}(p)\,  \bQ_{3}^{\ 1}
- \frac{1}{2} b(p) A_{2}^{\dagger}(p)\,  \bQ_{2}^{\dagger 4} \,, \non\\
\hat \bL_{2}^{\ 1}\, A_{4}^{\dagger}(p) &=&
A_{4}^{\dagger}(p)\, \hat \bL_{2}^{\ 1}
+ \frac{1}{2} d(p) A_{2}^{\dagger}(p)\,  \bQ_{4}^{\ 1}
+ \frac{1}{2} b(p) A_{2}^{\dagger}(p)\,  \bQ_{2}^{\dagger 3} \,.
\label{yangian1}
\ee 
We have verified that these relations, together with many others which
we have not listed here, are consistent with the bulk $S$-matrix
(\ref{bulkS3}), (\ref{bulkS4}), thereby confirming the $Y(su(2|2))$
Yangian symmetry of the latter.

\subsection{Boundary scattering}\label{subsec:fundamentalBound}

We consider now the problem of boundary scattering for the fundamental
excitations of open strings attached to $Y=0$ maximal giant graviton
branes \cite{HM}.
In order to describe boundary scattering, we extend (following
\cite{AN}) the bulk ZF algebra (\ref{bulkS1}) by introducing a
boundary operator ${\bf B}$ which creates the boundary-theory vacuum state
$|0\rangle_{B} = {\bf B}|0\rangle$  \cite{GZ}.  Since there is no
boundary degree of freedom for the $Y=0$ brane, the 
boundary operator is a scalar.
We define the fundamental (right) boundary $S$-matrix $R^{A}(p)$ by
\footnote{We consider here just a right boundary, since a left
boundary can be treated in a similar way \cite{AN}.}
\be
A^{\dagger}_{i}(p)\, {\bf B} = R_{\ i}^{A\, i'}(p)\, A^{\dagger}_{i'}(-p)\,  
{\bf B} \,.
\label{boundaryS}
\ee

As for the bulk, the boundary $S$-matrix can be determined 
from the symmetry of the problem. Indeed, the $Y=0$ brane preserves
only an $su(1|2)$ subalgebra \cite{HM}, which consists of the 
generators
\be
\bL_{1}^{\ 1}\,, \quad  \bL_{2}^{\ 2}\,, \quad\bH \,, \quad 
\bR_{\alpha}^{\ \beta}\,, \quad  \bQ_{\alpha}^{\ 1}\,, \quad  
\bQ_{1}^{\dagger \alpha} \quad \mbox{ with } 
\alpha\,, \beta \in \{3\,, 4\} \,.
\label{su12gens}
\ee
The vacuum state $|0\rangle_{B}$ is annihilated by 
each of these generators. Consider now one-particle states 
$A^{\dagger}_{i}(p)\, |0\rangle_{B}$.
Invariance under $\bL_{1}^{\ 1}$ and $\bR_{\alpha}^{\ \beta}$ implies that 
the boundary $S$-matrix is diagonal, with the structure
\be
R^{A}(p) = \diag(r_{1}\,, r_{2}\,,r \,,r ) \,.
\label{boundSform1}
\ee
Invariance under $\bQ_{3}^{\ 1}$ then determines the diagonal matrix 
elements,
\be
\frac{r_{1}}{r}&=&e^{-i p} \frac{a(p)}{a(-p)} =e^{-i p/2}\,, \non\\
\frac{r_{2}}{r}&=&e^{i p} \frac{b(-p)}{b(p)} =-e^{i p/2}\,.
\label{boundSresult1}
\ee
This result differs from the one in \cite{AN} due to the  
different expression (\ref{eta}) for $\eta(p)$.
This matrix satisfies the standard boundary Yang-Baxter equation
\be
\lefteqn{S_{12}^{AA}(p_{1}, p_{2})\, R_{1}^{A}(p_{1})\, 
S_{21}^{AA}(p_{2}, -p_{1})\, 
R_{2}^{A}(p_{2})} \non \\
& & = 
R_{2}^{A}(p_{2})\, S_{12}^{AA}(p_{1}, -p_{2})\, R_{1}^{A}(p_{1})\, 
S_{21}^{AA}(-p_{2}, -p_{1}) \,,
\label{BYBE}
\ee 
where 
\be
S_{21}^{AA}(p_{1}, p_{2}) 
= {\cal P}_{12}\, S_{12}^{AA}(p_{1}, p_{2})\, {\cal P}_{12}
= S_{12}^{AA}(p_{2}, p_{1})^{-1}
\,,
\ee
and ${\cal P}$ is the permutation matrix.

We now show that this boundary $S$-matrix also has a residual Yangian
symmetry.  Indeed, consider the charge $\tilde \bQ$ defined by
\be
\tilde \bQ = \hat \bL_{2}^{\ 1} + \frac{1}{2}\left( 
\bL_{2}^{\ 1} \bL_{1}^{\ 1} - \bL_{2}^{\ 1} \bL_{2}^{\ 2}
- \bQ_{2}^{\dagger 3} \bQ_{3}^{\ 1}
- \bQ_{2}^{\dagger 4} \bQ_{4}^{\ 1} \right) \,,
\label{tildeQ}
\ee
which we shall assume is also conserved.
This charge has the following action on the ZF operators,
\be
\tilde \bQ\,  A_{1}^{\dagger}(p) &=&
\left[-\frac{1}{2}i g u + \frac{1}{2} - a(p) d(p)\right] 
A_{2}^{\dagger}(p) + A_{2}^{\dagger}(p)\, (\bL_{1}^{\ 1}
- \bL_{2}^{\ 2} ) \non \\
& & + c(p) A_{4}^{\dagger}(p)\, \bQ_{3}^{\ 1}
- c(p) A_{3}^{\dagger}(p)\, \bQ_{4}^{\ 1} 
+ A_{1}^{\dagger}(p)\, \tilde \bQ \,, \non\\ 
\tilde \bQ\,  A_{2}^{\dagger}(p) &=&
A_{2}^{\dagger}(p)\, \tilde \bQ \,, \non\\ 
\tilde \bQ\,  A_{3}^{\dagger}(p) &=&
d(p) A_{2}^{\dagger}(p)\, \bQ_{3}^{\ 1}
+ A_{3}^{\dagger}(p)\, \tilde \bQ \,, \non\\ 
\tilde \bQ\,  A_{4}^{\dagger}(p) &=&
d(p) A_{2}^{\dagger}(p)\, \bQ_{4}^{\ 1}
+ A_{4}^{\dagger}(p)\, \tilde \bQ  \,,
\label{tildeQonZF}
\ee
as can be verified with the help of (\ref{yangian1}). The key 
point is that {\it all} the symmetry generators appearing on the RHS 
of (\ref{tildeQonZF}) annihilate the vacuum state. A similar 
construction has appeared in e.g. \cite{DMS, Do}.

In order to see the symmetry of the boundary $S$-matrix,
consider the action of $\tilde \bQ$ on the one-particle state
$A_{1}^{\dagger}(p) |0\rangle_{B}$.
Using (\ref{tildeQonZF}), we obtain
\be
\tilde \bQ\, A_{1}^{\dagger}(p) |0\rangle_{B} =
f(p) A_{2}^{\dagger}(p) |0\rangle_{B} 
= r_{2} f(p) A_{2}^{\dagger}(-p) |0\rangle_{B} \,,
\label{step1}
\ee
where in the first equality we have also introduced
\be
f(p) = -\frac{1}{2}i g u + \frac{1}{2} - a(p) d(p) =
-\frac{1}{2}\left[ i g u + a(p) d(p) + b(p) c(p) \right] \,,
\ee 
and we have used the fact that $\tilde \bQ$
and the $su(1|2)$ generators (\ref{su12gens}) annihilate the
vacuum state; and in the second equality we  have ``reflected'' using 
(\ref{boundaryS}), (\ref{boundSform1}) . Reversing the order of operations, 
we obtain
\be
\tilde \bQ\, A_{1}^{\dagger}(p) |0\rangle_{B} =
r_{1} \tilde \bQ\,  A_{1}^{\dagger}(-p) |0\rangle_{B} =
r_{1} f(-p) A_{2}^{\dagger}(-p) |0\rangle_{B} \,.
\label{step2}
\ee
Comparing (\ref{step1}) and  (\ref{step2}), we arrive at the relation 
\be
\frac{r_{1}}{r_{2}} = \frac{f(p)}{f(-p)} = -e^{-i p}  \,,
\ee
which is consistent with the result (\ref{boundSresult1}). 
We conclude that the charge $\tilde \bQ$ (\ref{tildeQ}) constructed
with the Yangian generator $\hat \bL_{2}^{\ 1}$ is a symmetry
of the boundary $S$-matrix. Additional charges of this sort can 
be constructed, but we shall not need them here.

\section{Two-particle bound state representation}\label{sec:boundstate}

We now proceed to determine the boundary $S$-matrix for two-particle 
bound states. We begin by reviewing some necessary results about
such states and their bulk scattering.

\subsection{Bulk scattering}\label{subsec:boundstateBulk}

The two-particle bound states form an 8-dimensional (atypical totally
symmetric) representation of $su(2|2)$.  Following the convenient
superspace formalism in \cite{AF}, the $su(2|2)$ generators can be
represented by differential operators on a vector space of analytic
functions of two bosonic variables $w_{a}$ and two fermionic variables
$\theta_{\alpha}$, as follows:
\be
\bL_{a}^{\ b} &=& 
w_{a}\frac{\partial}{\partial w_{b}}
-\frac{1}{2}\delta^{b}_{a}w_{c}\frac{\partial}{\partial w_{c}} \,, 
\qquad \qquad
\bR_{\alpha}^{\ \beta} = 
\theta_{\alpha}\frac{\partial}{\partial \theta_{\beta}}
-\frac{1}{2}\delta^{\beta}_{\alpha}\theta_{\gamma}
\frac{\partial}{\partial \theta_{\gamma}} \,, \non \\
\bQ_{\alpha}^{\ a} &=&
a\, \theta_{\alpha}\frac{\partial}{\partial w_{a}}
+ b\, \epsilon^{a b}\epsilon_{\alpha \beta} w_{b} 
\frac{\partial}{\partial \theta_{\beta}} \,, \qquad
\bQ_{a}^{\dagger \alpha} =
d\, w_{a} \frac{\partial}{\partial \theta_{\alpha}}
+ c\, \epsilon_{a b}\epsilon^{\alpha \beta}
\theta_{\beta}
\frac{\partial}{\partial w_{b}} \,, \non \\
\bC &=&
a b\, \left(w_{a}\frac{\partial}{\partial w_{a}}+
\theta_{\alpha}\frac{\partial}{\partial \theta_{\alpha}}\right)\,,
\qquad \quad
\bC^{\dagger} =
c d\, \left(w_{a}\frac{\partial}{\partial w_{a}}+
\theta_{\alpha}\frac{\partial}{\partial \theta_{\alpha}}\right)\,, \non \\
\bH &=&
(a d + b c) \left(w_{a}\frac{\partial}{\partial w_{a}}+
\theta_{\alpha}\frac{\partial}{\partial \theta_{\alpha}}\right)\,.
\label{superspacerep}
\ee 
The basis vectors $|e_{i}\rangle$ of the fundamental representation
are $|e_{a}\rangle = w_{a} \,, |e_{\alpha}\rangle = \theta_{\alpha}$;
and the basis vectors $|e_{J}\rangle$ of the two-particle bound state
representation are \cite{AF}
\be
|e_{1}\rangle &=& \frac{w_{1} w_{1}}{\sqrt2} \,, \qquad
|e_{2}\rangle = w_{1} w_{2} \,, \qquad 
|e_{3}\rangle = \frac{w_{2} w_{2}}{\sqrt2} \,, \qquad
|e_{4}\rangle = \theta_{3} \theta_{4} \,, \non \\
|e_{5}\rangle &=& w_{1} \theta_{3} \,, \qquad \
|e_{6}\rangle = w_{1} \theta_{4} \,, \qquad \
|e_{7}\rangle = w_{2} \theta_{3} \,, \qquad \
|e_{8}\rangle = w_{2} \theta_{4} \,.
\label{superspacebasis}
\ee 

We introduce corresponding ZF operators $B_{J}^{\dagger}(p)$, $J = 1, 
\ldots, 8$ for the two-particle bound states. 
The operators with $J = 1, \ldots, 4$ are bosonic, while the operators with
$J=5, \ldots, 8$ are fermionic.
The bulk $S$-matrix 
$S^{AB}(p_{1},p_{2})$ is defined by
\be
A_{i}^{\dagger}(p_{1})\, B_{J}^{\dagger}(p_{2}) = 
S_{\ \ \ \ i\, J}^{AB\ i' J'}(p_{1}, p_{2})\, 
B_{J'}^{\dagger}(p_{2})\, A_{i'}^{\dagger}(p_{1}) \,,
\label{bulkSAB}
\ee
and the bulk $S$-matrix 
$S^{BB}(p_{1},p_{2})$ is defined by
\be
B_{I}^{\dagger}(p_{1})\, B_{J}^{\dagger}(p_{2}) = 
S_{\ \ \ \ I\, J}^{BB\ I' J'}(p_{1}, p_{2})\, 
B_{J'}^{\dagger}(p_{2})\, B_{I'}^{\dagger}(p_{1}) \,.
\label{bulkSBB}
\ee
Both $S$-matrices are given in \cite{AF}.  \footnote{There are two
typos in the coefficients of $S^{AB}$ listed in Section 6.1.2 of
\cite{AF}.  In $a_{13}$, the factor in the numerator $(x_1^- - y_2^+)$
should be instead $(x_1^+ - y_2^+)$; i.e., the $x_1^-$ should be
changed to $x_1^+$.  And in $a_{14}$, the factor in the numerator
$(1-y_2^- x_1^-)$ should be instead $(1-y_2^- x_1^+)$ ; i.e., the
$x_1^-$ should be changed to $x_1^+$.}

The action of the bosonic $su(2|2)$ generators $\bJ \in 
\{ \bL_{a}^{\ b}\,, \bR_{\alpha}^{\ \beta} \}$ on these ZF operators 
is given by (cf. (\ref{repBulk1}))
\be
\bJ\, B_{J}^{\dagger}(p) = (\bJ)_{J}^{\ K} B_{K}^{\dagger}(p)
+ B_{J}^{\dagger}(p)\, \bJ \,,
\label{repBulkB1}
\ee 
and the action of the supersymmetry generators is given by (cf. 
(\ref{repBulk2}))
\be
\bQ_{\alpha}^{\ a}\, B_{J}^{\dagger}(p) &=& e^{-ip/2}\left[
(\bQ_{\alpha}^{\ a})_{J}^{\ K} B_{K}^{\dagger}(p)
+ (-1)^{\epsilon_{J}} B_{J}^{\dagger}(p)\, \bQ_{\alpha}^{\ a} 
\right]\,, \non \\
\bQ_{a}^{\dagger \alpha}\, B_{J}^{\dagger}(p) &=& e^{ip/2}\left[
(\bQ_{a}^{\dagger \alpha})_{J}^{\ K} B_{K}^{\dagger}(p)
+ (-1)^{\epsilon_{J}} B_{J}^{\dagger}(p)\, \bQ_{a}^{\dagger \alpha} 
\right]\,,
\label{repBulkB2}
\ee 
where the matrix elements $(\bJ)_{J}^{\ K}$, $(\bQ_{\alpha}^{\ 
a})_{J}^{\ K}$, $(\bQ_{a}^{\dagger \alpha})_{J}^{\ K} $ can be 
computed from (\ref{superspacerep}),  (\ref{superspacebasis}), 
and are provided for the reader's convenience in the Appendix;
$\epsilon_{J}$ is the Grassmann parity,
\be
\epsilon_{J}= \left\{ \begin{array}{cl} 
0 & \mbox{  for  } J=1,\ldots,4 \\
1 & \mbox{  for  } J=5,\ldots,8
\end{array}\right. \,,
\ee
and $a,b,c,d,x^{\pm}$ are given by (\ref{BulkParameters}),
(\ref{xpm}) with now $l=2$. Moreover, for the central charges,
\be
\bC\, B_{J}^{\dagger}(p) &=& e^{-i p}\left[
2a(p) b(p) B_{J}^{\dagger}(p) + 
B_{J}^{\dagger}(p)\, \bC \right]\,, \non \\
\bC^{\dagger}\, B_{J}^{\dagger}(p) &=& e^{i p}\left[ 
2c(p) d(p) B_{J}^{\dagger}(p) + 
B_{J}^{\dagger}(p)\, \bC^{\dagger} \right] \,, \non \\
\bH\, B_{J}^{\dagger}(p) &=& 2\left[a(p) d(p) + b(p) c(p)\right] B_{J}^{\dagger}(p) + 
B_{J}^{\dagger}(p)\, \bH \,.
\label{repBulkB3}
\ee 

The action of the $Y(su(2|2))$ Yangian generators on the ZF operators
can be inferred, similarly to the case of the fundamental
representation, from the coproducts given in \cite{Be2, AdLT} and the
relations (\ref{repBulkB1}), (\ref{repBulkB2}). In particular,
from the coproduct (\ref{coproduct}), we obtain
\be
\hat \bL_{2}^{\ 1}\, B_{1}^{\dagger}(p) &=& 
-\frac{\sqrt{2}}{2} i g u\, B_{2}^{\dagger}(p) 
+  B_{1}^{\dagger}(p)\, \hat \bL_{2}^{\ 1}
- B_{1}^{\dagger}(p)\,  \bL_{2}^{\ 1}
+ \frac{\sqrt{2}}{2} B_{2}^{\dagger}(p)\,  
(\bL_{1}^{\ 1} -  \bL_{2}^{\ 2}) \non\\
&+& \frac{\sqrt{2}}{2} c(p) B_{6}^{\dagger}(p)\,  \bQ_{3}^{\ 1}  
-\frac{\sqrt{2}}{2} c(p) B_{5}^{\dagger}(p)\,  \bQ_{4}^{\ 1}
-\frac{\sqrt{2}}{2} a(p) B_{5}^{\dagger}(p)\,  \bQ_{2}^{\dagger 3}
-\frac{\sqrt{2}}{2} a(p) B_{6}^{\dagger}(p)\,  \bQ_{2}^{\dagger 4} \,, \non\\
\hat \bL_{2}^{\ 1}\, B_{2}^{\dagger}(p) &=&
-\frac{\sqrt{2}}{2}i g u\, B_{3}^{\dagger}(p) + 
B_{2}^{\dagger}(p)\, \hat \bL_{2}^{\ 1}
+ \frac{\sqrt{2}}{2} B_{3}^{\dagger}(p)\,  (\bL_{1}^{\ 1} -  
\bL_{2}^{\ 2}) \non\\
&+&  \frac{1}{2} c(p) B_{8}^{\dagger}(p)\,  \bQ_{3}^{\ 1}  
-\frac{1}{2} c(p) B_{7}^{\dagger}(p)\,  \bQ_{4}^{\ 1}
-\frac{1}{2} a(p) B_{7}^{\dagger}(p)\,  \bQ_{2}^{\dagger 3}
-\frac{1}{2} a(p) B_{8}^{\dagger}(p)\,  \bQ_{2}^{\dagger 4} \,, \non\\
\hat \bL_{2}^{\ 1}\, B_{3}^{\dagger}(p) &=&
B_{3}^{\dagger}(p)\, \hat \bL_{2}^{\ 1}
+ B_{3}^{\dagger}(p)\, \bL_{2}^{\ 1} \,, \non\\
\hat \bL_{2}^{\ 1}\, B_{4}^{\dagger}(p) &=&
B_{4}^{\dagger}(p)\, \hat \bL_{2}^{\ 1} 
- \frac{1}{2} d(p) B_{8}^{\dagger}(p)\,  \bQ_{3}^{\ 1}  
+\frac{1}{2} d(p) B_{7}^{\dagger}(p)\,  \bQ_{4}^{\ 1}
+\frac{1}{2} b(p) B_{7}^{\dagger}(p)\,  \bQ_{2}^{\dagger 3} 
+\frac{1}{2} b(p) B_{8}^{\dagger}(p)\,  \bQ_{2}^{\dagger 4} \,, \non\\
\hat \bL_{2}^{\ 1}\, B_{5}^{\dagger}(p) &=& 
-\frac{1}{2}i g u\, B_{7}^{\dagger}(p) 
+  B_{5}^{\dagger}(p)\, \hat \bL_{2}^{\ 1}
- \frac{1}{2}B_{5}^{\dagger}(p)\,  \bL_{2}^{\ 1}
+ \frac{1}{2} B_{7}^{\dagger}(p)\,  
(\bL_{1}^{\ 1} -  \bL_{2}^{\ 2}) \non\\
&+& \frac{1}{2} \left[
d(p) B_{2}^{\dagger}(p) + c(p) B_{4}^{\dagger}(p) \right] \bQ_{3}^{\ 1} 
-\frac{1}{2} \left[
a(p) B_{4}^{\dagger}(p) + b(p) B_{2}^{\dagger}(p) \right] \bQ_{2}^{\dagger 4} \,, \non\\
\hat \bL_{2}^{\ 1}\, B_{6}^{\dagger}(p) &=& 
-\frac{1}{2}i g u\, B_{8}^{\dagger}(p) 
+  B_{6}^{\dagger}(p)\, \hat \bL_{2}^{\ 1}
- \frac{1}{2}B_{6}^{\dagger}(p)\,  \bL_{2}^{\ 1}
+ \frac{1}{2} B_{8}^{\dagger}(p)\,  
(\bL_{1}^{\ 1} -  \bL_{2}^{\ 2}) \non\\
&+& \frac{1}{2} \left[
d(p) B_{2}^{\dagger}(p) + c(p) B_{4}^{\dagger}(p) \right] \bQ_{4}^{\ 1} 
+\frac{1}{2} \left[
a(p) B_{4}^{\dagger}(p) + b(p) B_{2}^{\dagger}(p) 
\right] \bQ_{2}^{\dagger 3} \,, \non\\
\hat \bL_{2}^{\ 1}\, B_{7}^{\dagger}(p) &=&
B_{7}^{\dagger}(p)\, \hat \bL_{2}^{\ 1} 
+ \frac{1}{2}B_{7}^{\dagger}(p)\,  \bL_{2}^{\ 1}
+ \frac{\sqrt{2}}{2} d(p) B_{3}^{\dagger}(p)\,  \bQ_{3}^{\ 1}  
- \frac{\sqrt{2}}{2} b(p) B_{3}^{\dagger}(p)\,  \bQ_{2}^{\dagger 4} \,, \non\\
\hat \bL_{2}^{\ 1}\, B_{8}^{\dagger}(p) &=&
B_{8}^{\dagger}(p)\, \hat \bL_{2}^{\ 1} 
+ \frac{1}{2}B_{8}^{\dagger}(p)\,  \bL_{2}^{\ 1}
+ \frac{\sqrt{2}}{2} d(p) B_{3}^{\dagger}(p)\,  \bQ_{4}^{\ 1}  
+ \frac{\sqrt{2}}{2} b(p) B_{3}^{\dagger}(p)\,  \bQ_{2}^{\dagger 3} \,.
\label{yangian2}
\ee

\subsection{Boundary scattering}\label{subsec:boundstateBound}

We define the two-particle bound state boundary $S$-matrix $R^{B}(p)$ 
by
\be
B^{\dagger}_{J}(p)\, {\bf B} = R_{\ J}^{B\, J'}(p)\, 
B^{\dagger}_{J'}(-p)\,  
{\bf B} \,.
\label{boundarySB}
\ee
We assume, as in the case of the fundamental representation, that the 
$su(1|2)$ generators (\ref{su12gens}) annihilate the  vacuum state 
$|0\rangle_{B} = {\bf B} |0\rangle$. Consider now one-particle states 
$B^{\dagger}_{J}(p)\, |0\rangle_{B}$.
Invariance under $\bL_{1}^{\ 1}$ and $\bR_{\alpha}^{\ \beta}$ implies that 
the boundary $S$-matrix has the structure
\be
R^{B}(p) = \left( \begin{array}{cccccccc}
r_{1} \\
& r_{2} & & r_{5} \\
& & r_{3} \\
& r_{6} & & r_{4} \\
& & & & r_{7} \\
& & & & & r_{7} \\
& & & & & & r_{8} \\
& & & & & & & r_{8}
\end{array} \right) \,,
\label{boundSformB}
\ee
where matrix elements which are zero are left empty. Note that, 
in contrast with the $l=1$ (fundamental representation) case, the 
$l=2$ (two-particle bound state representation) boundary 
$S$-matrix is not diagonal.

Invariance under the supersymmetry generators $\bQ_{\alpha}^{\ 1}$, 
$\bQ_{1}^{\dagger \alpha}$ leads to the following set of linear equations,
\be
a(-p) e^{i p/2} r_{1} - a(p)  e^{-i p/2} r_{7} &=& 0 \,, \non \\
a(-p)  e^{i p/2} r_{2} - b(-p) e^{i p/2} r_{5} - a(p) e^{-i p/2}  r_{8} &=& 0 \,, \non \\
b(-p)  e^{i p/2} r_{4} - a(-p) e^{i p/2} r_{6} - b(p) e^{-i p/2} r_{8} &=& 0 \,, \non \\
a(p) e^{-i p/2} r_{4} + b(p) e^{-i p/2} r_{5} - a(-p) e^{i p/2} r_{7} &=& 0 \,, \non \\
b(p) e^{-i p/2} r_{2} + a(p) e^{-i p/2} r_{6}  - b(-p) e^{i p/2} r_{7} &=& 0 \,, \non \\
b(p) e^{-i p/2} r_{3} - b(-p)  e^{i p/2} r_{8} &=& 0 \,, 
\label{lineareqs1}
\ee
\be 
d(p)  e^{i p/2} r_{1} - d(-p) e^{-i p/2} r_{7} &=& 0 \,, \non \\
c(-p) e^{-i p/2} r_{2} + d(-p) e^{-i p/2} r_{5} - c(p)  e^{i p/2} r_{7} &=& 0 \,,\non \\
d(-p) e^{-i p/2} r_{4} + c(-p) e^{-i p/2} r_{6} - d(p)  e^{i 
p/2} r_{7} &=& 0 \,, \non \\
d(p)  e^{i p/2} r_{2} - c(p)  e^{i p/2} r_{6} - d(-p) e^{-i 
p/2} r_{8} &=& 0 \,, \non \\
c(p)  e^{i p/2} r_{4} - d(p)  e^{i p/2} r_{5} - c(-p) e^{-i 
p/2} r_{8} &=& 0 \,, \non \\
c(-p) e^{-i p/2} r_{3} - c(p)  e^{i p/2} r_{8} &=& 0 \,,
\label{lineareqs2}
\ee 
of which only 6 are independent.  Since there are 7 independent matrix
elements (any of the eight matrix elements can be set to unity, since
we are not concerned with the overall scalar factor), we conclude that
the ``ordinary'' $su(1|2)$ symmetry is not strong enough to determine
the boundary $S$-matrix.  A similar phenomenon was observed in
\cite{AF} for the bulk $S$-matrix $S^{BB}$.

We can obtain the needed additional linear equation by assuming
that the charge $\tilde \bQ$ (\ref{tildeQ}) is again conserved.
Indeed, using (\ref{yangian2}), we find that the action of this 
charge on the ZF operators is given by
\be
\tilde \bQ\, B_{1}^{\dagger}(p) &=& 
\sqrt{2}\left[ -\frac{1}{2} i g u +1-a(p) d(p)\right] \, B_{2}^{\dagger}(p) 
-\sqrt{2} a(p) c(p)\, B_{4}^{\dagger}(p)  
\non\\
&+& \sqrt{2} B_{2}^{\dagger}(p)\,  
(\bL_{1}^{\ 1} -  \bL_{2}^{\ 2}) + \sqrt{2} c(p) B_{6}^{\dagger}(p)\,  \bQ_{3}^{\ 1}  
-\sqrt{2} c(p) B_{5}^{\dagger}(p)\,  \bQ_{4}^{\ 1}
+ B_{1}^{\dagger}(p)\, \tilde \bQ \,, \non\\
\tilde \bQ\, B_{2}^{\dagger}(p) &=&
\sqrt{2}\left[ -\frac{1}{2}i g u - a(p) d(p)\right] B_{3}^{\dagger}(p) 
+ \sqrt{2} B_{3}^{\dagger}(p)\,  (\bL_{1}^{\ 1} -  
\bL_{2}^{\ 2}) \non\\
&+&   c(p) B_{8}^{\dagger}(p)\,  \bQ_{3}^{\ 1}  
- c(p) B_{7}^{\dagger}(p)\,  \bQ_{4}^{\ 1}
+ B_{2}^{\dagger}(p)\, \tilde\bQ \,, \non\\
\tilde \bQ\, B_{3}^{\dagger}(p) &=&
B_{3}^{\dagger}(p)\, \tilde \bQ \,, \non\\
\tilde \bQ\, B_{4}^{\dagger}(p) &=&
\sqrt{2} b(p) d(p)\,  B_{3}^{\dagger}(p)
- d(p) B_{8}^{\dagger}(p)\,  \bQ_{3}^{\ 1}  
+ d(p) B_{7}^{\dagger}(p)\,  \bQ_{4}^{\ 1}
+ B_{4}^{\dagger}(p)\, \tilde \bQ \,, \non\\
\tilde \bQ\, B_{5}^{\dagger}(p) &=& 
-\frac{1}{2}i g u\, B_{7}^{\dagger}(p) 
+  B_{7}^{\dagger}(p)\, (\bL_{1}^{\ 1} -  \bL_{2}^{\ 2})
+ \left[
d(p) B_{2}^{\dagger}(p) + c(p) B_{4}^{\dagger}(p) \right] \bQ_{3}^{\ 1} 
+  B_{5}^{\dagger}(p)\,\tilde \bQ \,, \non\\
\tilde \bQ\, B_{6}^{\dagger}(p) &=& 
-\frac{1}{2}i g u\, B_{8}^{\dagger}(p) 
+  B_{8}^{\dagger}(p)\,  
(\bL_{1}^{\ 1} -  \bL_{2}^{\ 2}) 
+ \left[
d(p) B_{2}^{\dagger}(p) + c(p) B_{4}^{\dagger}(p) \right] \bQ_{4}^{\ 1} 
+  B_{6}^{\dagger}(p)\,\tilde \bQ \,, \non\\
\tilde \bQ\, B_{7}^{\dagger}(p) &=&
\sqrt{2} d(p) B_{3}^{\dagger}(p)\,  \bQ_{3}^{\ 1}  
+ B_{7}^{\dagger}(p)\, \tilde \bQ \,, \non\\
\tilde \bQ\, B_{8}^{\dagger}(p) &=&
\sqrt{2} d(p) B_{3}^{\dagger}(p)\,  \bQ_{4}^{\ 1} 
+ B_{8}^{\dagger}(p)\, \tilde \bQ \,.
\label{tildeQonZF2}
\ee
Note that, as in the fundamental case (\ref{tildeQonZF}), all
the symmetry generators appearing on the RHS 
of (\ref{tildeQonZF2}) annihilate the vacuum state.
 
Consider the action of $\tilde \bQ$ on the one-particle state
$B_{5}^{\dagger}(p) |0\rangle_{B}$.
Using (\ref{tildeQonZF2}), we obtain
\be
\tilde \bQ\, B_{5}^{\dagger}(p) |0\rangle_{B} =
-\frac{1}{2}i g u\, B_{7}^{\dagger}(p)  |0\rangle_{B} 
= -r_{8} \frac{1}{2}i g u\, B_{7}^{\dagger}(-p) |0\rangle_{B} \,,
\label{step1B}
\ee
where in the first equality we have assumed that $\tilde \bQ$ and the
$su(1|2)$ generators (\ref{su12gens}) annihilate the vacuum state; and
in the second equality we have ``reflected'' using (\ref{boundarySB}),
(\ref{boundSformB}). Reversing the order of operations, we obtain
\be
\tilde \bQ\, B_{5}^{\dagger}(p) |0\rangle_{B} =
r_{7}\, \tilde \bQ\,  B_{5}^{\dagger}(-p) |0\rangle_{B} =
r_{7}\, \frac{1}{2}i g u\, B_{7}^{\dagger}(-p) |0\rangle_{B} \,.
\label{step2B}
\ee
Comparing (\ref{step1B}) and  (\ref{step2B}), we arrive at the 
desired relation \footnote{By acting with $\tilde \bQ$ on the
one-particle states $B_{1}^{\dagger}(p) |0\rangle_{B}$ and
$B_{2}^{\dagger}(p) |0\rangle_{B}$, one can derive in a similar manner
the following further relations,
\be
r_{1}\left[\frac{1}{2}i g u - b(-p) c(-p) \right]
+ r_{2}\left[\frac{1}{2}i g u + b(p) c(p) \right] + r_{6} a(p) c(p) 
&=& 0 \,, \non \\
r_{1} a(-p) c(-p) - r_{4} a(p) c(p)
- r_{5}\left[\frac{1}{2}i g u + b(p) c(p) \right] &=& 0 \,, \non \\
r_{2}\left[\frac{1}{2}i g u - a(-p) d(-p) \right] 
+ r_{3}\left[\frac{1}{2}i g u + a(p) d(p) \right] + r_{5} b(-p) d(-p) 
&=& 0 \,, \non 
\ee
which are also satisfied by the solution (\ref{boundSBresult}).}
\be
r_{7} = -r_{8}  \,.
\label{result}
\ee

Solving the linear equations (\ref{lineareqs1}),(\ref{lineareqs2}),
(\ref{result}), we obtain the following result for the boundary
$S$-matrix elements
\be
r_{1} &=& 1 \,, \qquad 
r_{2} = -\frac{\frac{1}{x^{-}}+x^{-}}{\frac{1}{x^{+}}+x^{-}} \,, \qquad
r_{3} = e^{i p} \,,  \qquad
r_{4} = \frac{\frac{1}{x^{+}}+x^{+}}{\frac{1}{x^{+}}+x^{-}}\,, \non \\
r_{5} &=& - r_{6} =  e^{i p/2} \frac{x^{-}-x^{+}}{1 + x^{-} x^{+}} \,, \qquad 
\quad \
r_{7} = - r_{8} = e^{i p/2} \,.
\label{boundSBresult}
\ee

The boundary $S$-matrix (\ref{boundSformB}), (\ref{boundSBresult}) and
its Yangian symmetry are our main results.  We have verified using
Mathematica that this boundary $S$-matrix $R^{B}(p)$ satisfies both
boundary Yang-Baxter equations
\be
\lefteqn{S_{12}^{AB}(p_{1}, p_{2})\, R_{1}^{A}(p_{1})\, 
S_{21}^{BA}(p_{2}, -p_{1})\, 
R_{2}^{B}(p_{2})} \non \\
& & = 
R_{2}^{B}(p_{2})\, S_{12}^{AB}(p_{1}, -p_{2})\, R_{1}^{A}(p_{1})\, 
S_{21}^{BA}(-p_{2}, -p_{1}) \,,
\label{BYBEAB}
\ee 
and
\be
\lefteqn{S_{12}^{BB}(p_{1}, p_{2})\, R_{1}^{B}(p_{1})\, 
S_{21}^{BB}(p_{2}, -p_{1})\, 
R_{2}^{B}(p_{2})} \non \\
& &  = 
R_{2}^{B}(p_{2})\, S_{12}^{BB}(p_{1}, -p_{2})\, R_{1}^{B}(p_{1})\, 
S_{21}^{BB}(-p_{2}, -p_{1}) \,,
\label{BYBEABB}
\ee 
where 
\be
S_{21}^{BA}(p_{1}, p_{2})  = S_{12}^{AB}(p_{2}, p_{1})^{-1} \,,
\ee
and $R^{A}(p)$ is given by (\ref{boundSform1}), (\ref{boundSresult1}).
We note that the boundary unitarity equation
\be
R^{B}(p)\, R^{B}(-p) = \id 
\label{boundunitarityR}
\ee 
is also satisfied. 

\section{Discussion}

We have showed that boundary scattering for the $Y=0$ brane has a
residual Yangian symmetry, which we have exploited to determine the
boundary $S$-matrix for two-particle bound states.  We expect that it
should be possible to further exploit this symmetry to determine the
boundary $S$-matrices for general $l$-particle bound states.  However,
this will require developing more powerful techniques, perhaps along
the lines of \cite{AdLT}.  We also expect that boundary scattering for
the $Z=0$ brane \cite{HM} has the full Yangian symmetry $Y(su(2|2))$,
which should determine the corresponding bound state boundary
$S$-matrices.  Moreover, the boundary $S$-matrices which have been
proposed for $D7$ and $D5$ branes \cite{CY} presumably also have
Yangian symmetry. We hope to be able to address these problems in 
the near future.

The fact that Yangian symmetry has been found in both bulk and
boundary scattering suggests that Yangian symmetry may be a generic
feature of AdS/CFT worldsheet scattering.  It would be interesting to
understand if there is any connection with the recently-discovered
Yangian symmetry in spacetime scattering \cite{DHP}, \cite{BHMP}.

\section*{Acknowledgments}

We are grateful to G. Arutyunov, S. Frolov and E. Quinn for 
valuable correspondence, and for making available to us their code
for the bulk $S$-matrix $S^{BB}$.
C.A. thanks the University of Miami for hospitality during the course
of this work.
This work was supported in part by KRF-2007-313-C00150
and WCU grant R32-2008-000-10130-0 (CA),
and by the National Science Foundation under Grants PHY-0554821 and
PHY-0854366 (RN).

\begin{appendix}
    
\section{Two-particle bound state representation of the $su(2|2)$ generators} 
\label{sec:l2rep}

We provide here the explicit two-particle bound state representation
of the $su(2|2)$ generators, which follows from (\ref{superspacerep}),
(\ref{superspacebasis}), and which is used for the computations in
Sec. \ref{sec:boundstate}.

For $\alpha = 3, 4$,

\parbox{4cm}{\begin{eqnarray*}
\bQ_{\alpha}^{\ 1}|e_{1}\rangle &=&\sqrt{2} a|e_{\alpha+2}\rangle \,,  \\
\bQ_{\alpha}^{\ 1}|e_{2}\rangle &=&a|e_{\alpha+4}\rangle \,,  \\
\bQ_{\alpha}^{\ 1}|e_{3}\rangle &=&0 \,, \\
\bQ_{\alpha}^{\ 1}|e_{4}\rangle &=&-b|e_{\alpha+4}\rangle \,,  \\
\bQ_{\alpha}^{\ 1}|e_{5}\rangle &=&- \delta_{\alpha}^{4}\left(b|e_{2}\rangle + a 
|e_{4}\rangle \right) \,,\\
\bQ_{\alpha}^{\ 1}|e_{6}\rangle &=&\delta_{\alpha}^{3} \left(b|e_{2}\rangle + a 
|e_{4}\rangle \right)  \,,  \\
\bQ_{\alpha}^{\ 1}|e_{7}\rangle &=&-\delta_{\alpha}^{4}\sqrt{2} b |e_{3}\rangle \,,  \\
\bQ_{\alpha}^{\ 1}|e_{8}\rangle &=&\delta_{\alpha}^{3}\sqrt{2} b 
|e_{3}\rangle \,, 
\end{eqnarray*}}
\hfill \parbox{8cm}{\begin{eqnarray}
\bQ_{\alpha}^{\ 2}|e_{1}\rangle &=&0 \,, \non \\
\bQ_{\alpha}^{\ 2}|e_{2}\rangle &=&a|e_{\alpha+2}\rangle \,, \non \\
\bQ_{\alpha}^{\ 2}|e_{3}\rangle &=&\sqrt{2}a|e_{\alpha+4}\rangle \,, \non \\
\bQ_{\alpha}^{\ 2}|e_{4}\rangle &=&b|e_{\alpha+2}\rangle \,, \non \\
\bQ_{\alpha}^{\ 2}|e_{5}\rangle &=&\delta_{\alpha}^{4} \sqrt{2}b|e_{1}\rangle\,, \non \\
\bQ_{\alpha}^{\ 2}|e_{6}\rangle &=&-\delta_{\alpha}^{3}\sqrt{2}b|e_{1}\rangle \,, \non \\
\bQ_{\alpha}^{\ 2}|e_{7}\rangle &=&\delta_{\alpha}^{4} 
\left(b|e_{2}\rangle - a 
|e_{4}\rangle \right)  \,, \non \\
\bQ_{\alpha}^{\ 2}|e_{8}\rangle &=&\delta_{\alpha}^{3} \left(-b|e_{2}\rangle + a 
|e_{4}\rangle \right)  \,,
\end{eqnarray}}

and

\parbox{4cm}{\begin{eqnarray*}
\bQ_{1}^{\dagger \alpha}|e_{1}\rangle &=&0 \,,  \\
\bQ_{1}^{\dagger \alpha}|e_{2}\rangle &=&c \epsilon^{\alpha 
\beta}|e_{\beta+2}\rangle \,,  \\
\bQ_{1}^{\dagger \alpha}|e_{3}\rangle &=&\sqrt{2}c \epsilon^{\alpha 
\beta}|e_{\beta+4}\rangle \,,  \\
\bQ_{1}^{\dagger \alpha}|e_{4}\rangle &=&d \epsilon^{\alpha 
\beta}|e_{\beta+2}\rangle \,,  \\
\bQ_{1}^{\dagger \alpha}|e_{5}\rangle &=&\delta_{3}^{\alpha} \sqrt{2} 
d |e_{1}\rangle \,,\\
\bQ_{1}^{\dagger \alpha}|e_{6}\rangle &=& \delta_{4}^{\alpha} \sqrt{2} 
d |e_{1}\rangle \,,  \\
\bQ_{1}^{\dagger \alpha}|e_{7}\rangle &=&\delta_{3}^{\alpha} 
\left(d|e_{2}\rangle - c |e_{4}\rangle \right) \,,  \\
\bQ_{1}^{\dagger \alpha}|e_{8}\rangle &=&\delta_{4}^{\alpha} 
\left(d|e_{2}\rangle - c |e_{4}\rangle \right) \,, 
\end{eqnarray*}}
\hfill \parbox{8cm}
{\begin{eqnarray}
\bQ_{2}^{\dagger \alpha}|e_{1}\rangle &=&-\sqrt{2}c \epsilon^{\alpha 
\beta}|e_{\beta+2}\rangle \,, \non  \\
\bQ_{2}^{\dagger \alpha}|e_{2}\rangle &=& -c \epsilon^{\alpha 
\beta}|e_{\beta+4}\rangle\,, \non  \\
\bQ_{2}^{\dagger \alpha}|e_{3}\rangle &=&0\,, \non  \\
\bQ_{2}^{\dagger \alpha}|e_{4}\rangle &=&d \epsilon^{\alpha 
\beta}|e_{\beta+4}\rangle \,,  \non \\
\bQ_{2}^{\dagger \alpha}|e_{5}\rangle &=&\delta_{3}^{\alpha} 
\left(d|e_{2}\rangle + c |e_{4}\rangle \right) \,,\non \\
\bQ_{2}^{\dagger \alpha}|e_{6}\rangle &=& \delta_{4}^{\alpha} 
\left(d|e_{2}\rangle + c |e_{4}\rangle \right) \,, \non  \\
\bQ_{2}^{\dagger \alpha}|e_{7}\rangle &=& \delta_{3}^{\alpha} 
\sqrt{2}d |e_{3}\rangle\,, \non  \\
\bQ_{2}^{\dagger \alpha}|e_{8}\rangle &=&\delta_{4}^{\alpha} 
\sqrt{2}d |e_{3}\rangle \,. 
\end{eqnarray}}

Moreover,

\parbox{4cm}{\begin{eqnarray*}
\bL_{1}^{\ 2}|e_{1}\rangle &=&0 \,,  \\
\bL_{1}^{\ 2}|e_{2}\rangle &=&\sqrt{2}|e_{1}\rangle \,,  \\
\bL_{1}^{\ 2}|e_{3}\rangle &=&\sqrt{2}|e_{2}\rangle \,,  \\
\bL_{1}^{\ 2}|e_{4}\rangle &=&0 \,,  \\
\bL_{1}^{\ 2}|e_{5}\rangle &=&0 \,,  \\
\bL_{1}^{\ 2}|e_{6}\rangle &=&0 \,,  \\
\bL_{1}^{\ 2}|e_{7}\rangle &=&|e_{5}\rangle \,,  \\
\bL_{1}^{\ 2}|e_{8}\rangle &=&|e_{6}\rangle \,,  
\end{eqnarray*}}
\hfill \parbox{8cm}
{\begin{eqnarray}
\bL_{2}^{\ 1}|e_{1}\rangle &=&\sqrt{2}|e_{2}\rangle \,, \non \\
\bL_{2}^{\ 1}|e_{2}\rangle &=&\sqrt{2}|e_{3}\rangle \,,  \non \\
\bL_{2}^{\ 1}|e_{3}\rangle &=&0 \,, \non  \\
\bL_{2}^{\ 1}|e_{4}\rangle &=&0 \,,  \non \\
\bL_{2}^{\ 1}|e_{5}\rangle &=&|e_{7}\rangle \,,  \non \\
\bL_{2}^{\ 1}|e_{6}\rangle &=&|e_{8}\rangle \,,  \non \\
\bL_{2}^{\ 1}|e_{7}\rangle &=&0 \,,  \non \\
\bL_{2}^{\ 1}|e_{8}\rangle &=&0 \,,  
\end{eqnarray}}

and $\bR_{3}^{\ 4}|e_{J}\rangle = 0 = \bR_{4}^{\ 3}|e_{J}\rangle$ 
except for the following:
\be
\bR_{3}^{\ 4}|e_{6}\rangle &=& |e_{5}\rangle \,, \qquad 
\bR_{3}^{\ 4}|e_{8}\rangle = |e_{7}\rangle \,, \non \\
\bR_{4}^{\ 3}|e_{5}\rangle &=& |e_{6}\rangle \,, \qquad 
\bR_{4}^{\ 3}|e_{7}\rangle = |e_{8}\rangle \,.
\ee 

Finally, $\bL_{1}^{\ 1}\,, \bL_{2}^{\ 2}\,, \bR_{3}^{\ 3}\,, 
\bR_{4}^{\ 4}$ are represented by diagonal matrices, whose 
eigenvalues are given in the following table:

\begin{table}[h]
    \centering
    \begin{tabular}{|r|r|r|r|r|}\hline
	& $\bL_{1}^{\ 1}$ & $\bL_{2}^{\ 2}$ & $\bR_{3}^{\ 3}$  & 
	$\bR_{4}^{\ 4}$ \\
	\hline
$|e_{1}\rangle$	& 1 & -1 & 0 & 0\\
$|e_{2}\rangle$   & 0 &  0 & 0 & 0\\
$|e_{3}\rangle$	& -1 & 1 & 0 & 0\\
$|e_{4}\rangle$   & 0 &  0 & 0 & 0\\
$|e_{5}\rangle$  & $\frac{1}{2}$  & $-\frac{1}{2}$ & $\frac{1}{2}$ & 
$-\frac{1}{2}$\\
$|e_{6}\rangle$   & $\frac{1}{2}$  & $-\frac{1}{2}$ & $-\frac{1}{2}$ & $\frac{1}{2}$\\
$|e_{7}\rangle$   & $-\frac{1}{2}$  & $\frac{1}{2}$ & $\frac{1}{2}$ 
& $-\frac{1}{2}$\\
$|e_{8}\rangle$   & $-\frac{1}{2}$  & $\frac{1}{2}$ & $-\frac{1}{2}$ & $\frac{1}{2}$\\
\hline
   \end{tabular}
   \end{table}

\end{appendix}

\end{document}